\documentclass[a4paper,twocolumn,secnumarabic,amssymb, nobibnotes, aps, prd]{revtex4-2}

\usepackage{CJK,upgreek,fancyhdr}
\usepackage{graphicx}
\usepackage{booktabs}
\usepackage{amssymb,bm,mathrsfs,bbm,amscd}
\usepackage[tbtags]{amsmath}
\usepackage{lastpage}
\usepackage{float}
\usepackage{tabularx}

\def\braketm#1#2#3{\Bigl\langle#1\Bigr|#2\Bigr|#3\Bigr\rangle}

\def\3F2#1#2#3{_3\mathrm{F}_2 \left( #1;#2 | #3\right)}
\def\binomial#1#2{\begin{pmatrix}#1\\#2 \end{pmatrix}}

\begin{document}



\title{Energy corrections due to the noncommutative phase-space of the charged isotropic harmonic oscillator in a uniform magnetic field in 3D }

\author{Muhittin Cenk Eser}%
\email{muhittin.eser@emu.edu.tr}
\affiliation{Department of Physics, Eastern Mediterranean University, 99628 Famagusta, North Cyprus}

\author{Mustafa Riza}%
\email{mustafa.riza@emu.edu.tr}
\affiliation{Department of Physics, Eastern Mediterranean University, 99628 Famagusta, North Cyprus}
\date{\today}

\begin{abstract}
In this study, we investigate the effects of noncommutative Quantum Mechanics in three dimensions on the energy levels of a charged isotropic harmonic oscillator in the presence of a uniform magnetic field in the $z$-direction. The extension of this problem to three dimensions proves to be non-trivial. We obtain the first-order corrections to the energy-levels in closed form in the low energy limit of weak noncommutativity. The most important result we can note is that all energy corrections due to noncommutativity are negative and their magnitude increase with increasing Quantum numbers and magnetic field. 
\end{abstract}

\maketitle
\section{Introduction}

With Heisenberg's introduction of the Uncertainty Principle \cite{heisenberg1985anschaulichen}, the classical paradigm that position and momentum commutate at no cost is crushed. Furthermore, the discussion of a charged particle in an electromagnetic field in the framework of Quantum Mechanics inevitably leads to the introduction of the kinetic momentum operator, which in contrast to the canonical momentum operator, does not commute. The noncommutativity of the kinetic momentum operator indicates that the magnetic field's presence modifies momentum space. These two facts, emerging from  Quantum Mechanics' nature, evidently bring up the question, is the assumption of the commutation of the position and momentum operators among themselves an accurate assumption? Or, under which conditions is the assumption of the vanishing commutators of $[x_i,x_j]=0$ and $[p_i,p_j]=0$ correct? The field dedicated to the study of non vanishing position and momentum commutators is called Noncommutative Quantum Mechanics. It is evident that in the energy domain of textbook quantum mechanics, the commutators in position and momentum are vanishing. Once the energy is pushed closer towards the Planck Energy, the  effects of noncommutativity of the momentum and position operators can be observed \cite{DEY2017578}. 

In his pioneering work, Hartland Snyder \cite{snyder1947quantized} noticed  that in Field Theory Lorenz invariance  does not necessarily require noncommutativity of the position and momentum operators. Snyder's work \cite{snyder1947quantized} lead to the detailed discussion of the Quantum Field Theory in noncommutative spaces by \citeauthor{SZABO2003207}  \cite{SZABO2003207} and \citeauthor{seiberg1999string}  \cite{seiberg1999string}. One of the first formulations of non-relativistic Quantum Mechanics in noncommutative space was presented by \citeauthor{chaturvedi1993non} in \cite{chaturvedi1993non}. Based on these ideas, Noncommutative Quantum Mechanics was proposed by \citeauthor{RN78} in \cite{RN78}. Noncommutativity is generally associated with the effect of the geometry of the space \cite{connes1998noncommutative,PhysRevD62066004}.  The Klein-Gordon, the  Schr\"odinger, and Pauli-Dirac oscillators in noncommutative phase-space have been studied by \citeauthor{jian2008klein} in \cite{jian2008klein} and \citeauthor{santos2011schroedinger} in\cite{santos2011schroedinger}. Furthermore, more fundamental problems like the Bohr-van-Leeuwen theorem \cite{biswas2017bohr},  stating explicitly that  magnetization is a purely Quantum Mechanical effect, is discussed in the framework of Noncommutative Quantum Mechanics.

Moreover, the noncommutative geometry in space seems to be a reasonable approach to the limitation of the position uncertainty leading us to the General Uncertainty Principle discussed e.g., by \citeauthor{PhysRevD.52.1108} in \cite{PhysRevD.52.1108}, \citeauthor{doi:10.1139/cjp-2015-0456} in \cite{doi:10.1139/cjp-2015-0456}, or \citeauthor{PhysRevD.96.066008} \cite{PhysRevD.96.066008}.  \citeauthor{dey2012symmetric} \cite{dey2012symmetric} showed explicitly that noncommutativity of the phase space gives rise to minimum length and minimum momentum uncertainties. \citeauthor{DEY2017578} \cite{DEY2017578} also stated that the minimum length  from noncommutativity is consistent with the General Uncertainty Principle in \cite{PhysRevD.52.1108}, yielding to $\left(\Delta x_i\right)_{min} = l_p \frac{\hbar}{2} \sqrt{\eta \theta}$, where $l_p=\sqrt{\hbar G/c^3} \sim  10^{-35}m$ denotes the Planck-Length. Based on the experimental bound for $\theta$ the minimal length is of the order of $10^6 l_p$. Based on the experimental setup proposed in \cite{DEY2017578} the Planck Length as minimal length can be reached and effects of Quantum Gravity can be observed. Furthermore, the idea of minimum length for resolving the UV singularity motivated noncommutative space-time in Quantum Field Theory. Studies in String Theory \cite{gross1988string,AHARONY2000183,magueijo2005string} and in Loop Quantum Gravity \cite{rovelli2011simple,rovelli2011new,smolin2004invitation} support this idea. However, only the formation of a black hole provides the necessary conditions for arbitrarily high precision in the position\cite{doplicher1995quantum, Bronstein:2012wo}. Consequently, the physical limitation on the shortest distance leads to a UV cutoff \cite{kurkov2014spectral}. Moreover, the noncommutative phase-space and its space-time symmetry in $2+1$ dimensions have been discussed by \citeauthor{kang2010non} \cite{kang2010non}.
 
The relationship between the General Uncertainty Principle as proposed in \cite{PhysRevD.96.066008,PhysRevD.52.1108} and noncommutative Quantum Mechanics needs to be analyzed in detail, as both approaches lead to the concept of minimal length uncertainty primarily, and minimal momentum uncertainty in the second instance. A minimal uncertainty in length and momentum leads also to an important conclusion in information theory, namely that the total information in the universe is bounded. This will form a reasonable extension to \citeauthor{PhysRevD.23.287}'s work \cite{PhysRevD.23.287}.

This study is dedicated to the noncommutative 3D isotropic harmonic oscillator in a homogeneous magnetic field. The application of the magnetic field to the isotropic harmonic oscillator turns the isotropic harmonic oscillator into an anisotropic harmonic oscillator.  The anisotropic harmonic oscillator has a wide range of applications in mathematical physics, Quantum theory, and condensed matter physics commutative as well as noncommutative. In the commutative case, we can find a wide field of applications in the literature. e.g., Petreska has applied the concept of the anisotropic harmonic oscillator to different problems in Quantum Physics \cite{ISI:000340928400019,ISI:000330500800003,Petreska:2013tx,ISI:000280141900026,ISI:000245480400031}. On the other hand, it serves also as a perfect model in the discussion of Quantum dots in condensed matter physics \cite{ISI:000459579900004,ISI:000314075100007,ISI:000296002200016,ISI:000291887600018,ISI:000274002500073,ISI:000245329600118,ISI:000244533800037,ISI:000243701700006,ISI:000239426800091,ISI:000232075600004,ISI:000185420400015} and atomic physics \cite{ISI:000551735100001,ISI:000295083400003,ISI:000275072500079,ISI:000234120800032,ISI:000231564200081,ISI:000221428300015,ISI:000220636000015,ISI:000185966200005,ISI:000184012700003,ISI:000184012400015}. In the noncommutative case, the discussions  are mainly carried out in the noncommutative plane, i.e. noncommutativity is only employed to the $xy$-plane for both the position and momentum. With respect to this  \citeauthor{gao2008exact} solve the isotropic charged harmonic oscillator in a uniform magnetic field 2D Noncommutative Quantum Mechanics \cite{gao2008exact}. The isotropic harmonic oscillator in a constant magnetic field is a subset of the anisotropic harmonic oscillator. The anisotropic harmonic oscillator was also discussed under different aspects in the framework of noncommutative Quantum Mechanics \cite{ISI:000607074300004,ISI:000502888700037,nath2017noncommutative} explicitly. \citeauthor{ISI:000607074300004} show in \cite{ISI:000607074300004} that entanglement induces noncommutativity in space in the example of the anisotropic harmonic oscillator. Furthermore, \citeauthor{ISI:000502888700037} discuss the impact of noncommutativity on the uncertainty and the Shanon entropy for the 2D anisotropic harmonic oscillator in presence of a magnetic field \cite{ISI:000502888700037}. The 2D noncommutative anisotropic harmonic oscillator in a homogeneous magnetic field has been discussed by \citeauthor{nath2017noncommutative} in \cite{nath2017noncommutative}. In contrast to the studies cited, this study illuminates the energy corrections due to 3D noncommutativity as a function of the magnetic field in the low energy limit according to \cite{harko2019energy}. Additionally,various publications are dedicated to the charged Quantum harmonic oscillator in the presence of a constant or time-varying electromagnetic field in noncommutative Quantum Mechanics e.g.,\cite{liang2010time}. Moreover, the magnetic field's impact on noncommutativity has been discussed in numerous works, especially in the context of the Landau problem \cite{mamat2016,RN51,RN96,RN36,RN81,RN97,RN84,RN41,RN34,RN77,RN68,iengo2002landau,gangopadhyay2015landau}. There are several discussions on the noncommutative Quantum Hall effect \cite{RN40,RN101,RN29,RN33} as well.  The minimally coupled charged harmonic oscillator to the magnetic  field in a noncommutative plane has been studied extensively by \citeauthor{jing2009non} \cite{jing2009non}.
Some more mathematical discussions on the  noncommutativity of Quantum Mechanics can be found e.g., in \cite{chakraborty2010twist,kuznetsova2013effects,banerjee2002novel}.
Finally,  \citeauthor{hassanabadi2014dirac} studied the Dirac oscillator in the presence of the Aharonov-Bohm effect in noncommutative and commutative spaces \cite{hassanabadi2014dirac}.

The fact that the magnetic field modifies the momentum space leading to the noncommutativity  of the kinetic momentum operator  on the one hand, and various studies related to the General Uncertainty Principle, backed also by String theory, suggest that the existence of a minimal length on the other hand, support the approach in Noncommutative Quantum Mechanics including the noncommutativity of the position and the momentum operators. The noncommutativity of the position and momentum operators  indicate a minimum length and a minimum momentum. Continuing this train of thought will lead to the conclusion that all physical quantities are quantized and have a minimum size.

 Throughout this manuscript we will denote  $\hat{x}_i$ as the noncommutative position operator and $\hat{p}_i$ as the noncommutative momentum operator in contrast to the standard position operator $x_i$ and the standard momentum operator $p_i$. The basic properties of noncommutative phase-space according to e.g., \citeauthor{RN78} \cite{RN78} stating the commutator relationships of the noncommutative position operators and the noncommutative momentum operators as:

\begin{equation}
	\left[\hat{x}_i,\hat{x}_j\right]=i \theta_{ij}, \quad \left[\hat{p}_i,\hat{p}_j\right]= i \eta_{ij}
	\label{eqn:1}
\end{equation}
where $\theta_{ij}$ and $\eta_{ij}$ are both  antisymmetric tensors. For further reading on antisymmetric tensors, we refer to \cite{hess2015tensors}. 

Consequently, as one can verify easily, the relationship between noncommutative operators $\hat{x}_i$ and $\hat{p}_i$ with their commutative counterparts can be written as

\begin{eqnarray}
\hat{x}_i &=& \alpha x_i - \frac{1}{2 \alpha \hbar} \theta_{ij} p_j \label{eq:xnc}\\
\hat{p}_i &=& \alpha p_i + \frac{1}{2 \alpha \hbar} \eta_{ij} x_j,  \label{eq:pnc}
\end{eqnarray}

where $\alpha \in (0,1)$ is the scaling constant related to the noncommutativity of the phase-space and $\eta_{ij}$,  and $\theta_{ij}$ are antisymmetric tensors. So, generally, we can express the tensors $\eta_{ij}$ and $\theta_{ij}$ as following:
\begin{eqnarray}
\eta_{ij} &=& \eta \lambda_{ij}, \label{eq:etaij}\\
\theta_{ij} &=& \theta \lambda_{ij}\label{eq:thetaij},
\end{eqnarray}
where $\lambda_{ij}$ denotes an antisymmetric tensor \cite{hess2015tensors}.

Mathematically, the noncommutativity of the base manifold can be realized by application the Weyl-Moyal star product \cite{mezincescu2000star}
\begin{multline}
\left(f \star g\right)(x,p) = e^{i\frac{1}{2\alpha^2} \theta_{ij}\partial_i^x \partial_j^x+ i\frac{1}{2\alpha^2} \eta_{ij}\partial_i^p \partial_j^p} f(x) g(y) =\\ =f(x,p) g(x,p) + \frac{i \theta_{ij}}{2\alpha^2} \partial_i^x f \partial_j^x g  \Biggr|_{x_i = x_j} + \frac{i \eta_{ij}}{2\alpha^2} \partial_i^p f \partial_j^p g  \Biggr|_{p_i = p_j}+ \\+ {\cal{O}}\left(\theta_{ij}^2\right)+  { \cal{O}}\left(\eta_{ij}^2\right)+{ \cal{O}}\left(\theta_{ij}\eta_{ij} \right) \label{eq:weyl-moyal}
\end{multline}
So, the shift from ordinary Quantum Mechanics to Noncommutative Quantum Mechanics is performed by employing the Weyl-Moyal product  \eqref{eq:weyl-moyal} instead of the ordinary product.  Hence, the Noncommutative Time-Independent Schr\"odinger Equation becomes
\begin{equation}
H(x,p) \star \psi(x) = E \psi(x).
\end{equation}
By employing the  Bopp's shift \cite{curtright1998features}, we can turn the Weyl-Moyal product again to the ordinary product by substituting $x$ and $p$ in the noncommutative equation by $\hat{x}$ and $\hat{p}$, namely 
\begin{equation}
H(x,p) \star \psi(x)  = H(\hat{x}, \hat{p}) \psi(x).\label{eq:wm-normal}
\end{equation}

 \citeauthor{harko2019energy} \cite{harko2019energy} state that the noncommutativity parameters $\eta$ and $\theta$ can be considered as energy-dependent and that both become sufficiently small in the low energy limit. Employing this fact, gives the justification of the possibility of the application of perturbation theory in the low energy limit.

In light of this, we will discuss the noncommutative charged harmonic oscillator in the presence of a uniform magnetic field employing noncommutativity to all three spacial parameters by including also the $z$-direction into the noncommutative framework. Therefore, first we will discuss the change to the noncommutative algebra by considering the commutator $[\hat{x}_i, \hat{p}_j]$ in the noncommutative plane and space in section \ref{sec:commutator}.  In the next section,\label{sec:2} we will discuss the noncommutative Hamiltonian of the charged particle in a 3D isotropic harmonic oscillator in the presence of a uniform magnetic field where we will expand the Hamiltonian in terms of $\theta$ and $\eta$. As this Hamiltonian proves to be non-trivial, the corrections to the eigenenergies due to the magnitude of the magnetic field will be calculated in section \ref{sec:perturbation} in first-order perturbation theory in $\eta$ and $\theta$, i.e. in the domain of weak noncommutativity in the low energy limit.  Finally, we will carry out a short analysis of the corrections of the eigenenergies in section \ref{sec:discussion} on the dependence of the energy corrections on the magnitude of the magnetic field for different values of the Quantum numbers and close this study with some concluding remarks.

\section{The commutator $[\hat{x}_i, \hat{p}_j]$ in the noncommutative plane and space}

\label{sec:commutator}

For completeness, let us recall the commutators $[\hat{x}_i, \hat{x}_j]$ and $[\hat{p}_i,\hat{p}_j]$.

\begin{equation}
	\left[\hat{x}_i,\hat{x}_j\right]=i \theta_{ij}, \quad \left[\hat{p}_i,\hat{p}_j\right]= i \eta_{ij}
	\label{eqn:1}
\end{equation}
where $\theta_{ij}$ and $\eta_{ij}$ are both  antisymmetric tensors. 

Yielding to  the relationship between noncommutative operators $\hat{x}_i$ and $\hat{p}_i$ with their commutative counterparts

\begin{eqnarray}
\hat{x}_i &=& \alpha x_i - \frac{1}{2 \alpha \hbar} \theta_{ij} p_j \nonumber\\
\hat{p}_i &=& \alpha p_i + \frac{1}{2 \alpha \hbar} \eta_{ij} x_j,  \nonumber
\end{eqnarray}

where $\alpha \in (0,1)$ is the scaling constant related to the noncommutativity of the phase-space  and $\eta_{ij}$,  and $\theta_{ij}$ are antisymmetric tensors. So, generally, we can express the tensors $\eta_{ij}$ and $\theta_{ij}$ as following:
\begin{eqnarray*}
\eta_{ij} &=& \eta \lambda_{ij}, \\
\theta_{ij} &=& \theta \lambda_{ij},
\end{eqnarray*}
where $\lambda_{ij}$ denotes an antisymmetric tensor.

The difference between the noncommutative plane and space is manifested in the definition of the antisymmetric tensor $\lambda_{ij}$. 
In the noncommutative plane the antisymmetric tensor $\lambda_{ij}$ is given as:

\begin{equation}
\lambda_{ij} =
\begin{cases}
1 & \text{if } ij= 12\\
-1 & \text{if } ij= 21\\
0 & \text{else }
\end{cases} \qquad .\label{eq:epsilonij2D}
\end{equation}

By extending the discussion to the 3D noncommutative space, a redefinition of the epsilon tensor is needed $\tilde{\lambda}_{ij}$ is defined as
\begin{equation}
\tilde{\lambda}_{ij} =
\begin{cases}
1 & \text{if } ij= 12,23,31\\
-1 & \text{if } ij= 21,32,13\\
0 & \text{else }
\end{cases} \qquad .\label{eq:epsilonij3D}
\end{equation}

Let us first discuss the impact of the extension of the antisymmetric tensor from the noncommutative plane $\lambda_{ij}$ to the noncommutative space  $\tilde{\lambda}_{ij}$ on the commutator $[\hat{x}_i, \hat{p}_j]$. The commutator of the noncommutative position and momentum operators can be calculated straight forward independent of the noncommutativity covering only the plane or the whole space
\begin{multline}
\left[ \hat{x}_i,\hat{p}_j\right] = \left[\alpha x_i - \frac{1}{2 \alpha \hbar} \theta_{ij} p_j ,\alpha p_i + \frac{1}{2 \alpha \hbar} \eta_{ij} x_j\right]=\\= i \hbar\alpha^2\delta_{ij} + i\frac{\theta \eta}{4 \alpha^2 \hbar} \lambda_{i\mu} \lambda_{j\mu}.\label{eq:ncxpcommutator}
\end{multline}
For the 3D noncommutative space $\lambda_{i\mu} \lambda_{j\mu}$ is substituted by $\tilde{\lambda}_{i\mu} \tilde{\lambda}_{j\mu}$.

The difference between the two cases of the noncommutative plane and the noncommutative space is manifested in the product of the antisymmetric tensors $\lambda_{i\mu}\lambda_{j\mu}$ and $\tilde{\lambda}_{i\mu} \tilde{\lambda}_{j\mu}$. Using the properties of the $\lambda$ tensor \eqref{eq:epsilonij2D}  for the noncommutative plane, we get for this product
\begin{equation}
\lambda_{i\mu} \lambda_{j\mu} = -\delta_{ij}, \label{eq:eimejmplane}
\end{equation}
where $\delta_{ij}$ denotes the Kronecker-$\delta$. Whereas the product of the two $\lambda$ tensors \eqref{eq:epsilonij3D} in the noncommutative space (3D) is 
\begin{equation}
\tilde{\lambda}_{i\mu} \tilde{\lambda}_{j\mu} = -3\delta_{ij}+1.\label{eq:eimejmspace}
\end{equation}

With \eqref{eq:eimejmplane} we get for the commutator \eqref{eq:ncxpcommutator} in the noncommutative plane
\begin{equation}
\left[ \hat{x}_i,\hat{p}_j\right] =i \hbar\alpha^2\delta_{ij} - i\frac{\theta \eta}{4 \alpha^2 \hbar}\delta_{ij},\label{eq:nccomm2D}
\end{equation}
and with \eqref{eq:eimejmspace} we get for the commutator  \eqref{eq:ncxpcommutator}  in the noncommutative space
\begin{equation}
\left[ \hat{x}_i,\hat{p}_j\right] =i \hbar\alpha^2\delta_{ij} - i\frac{\theta \eta}{4 \alpha^2 \hbar}\left(3\delta_{ij}-1\right).\label{eq:nccomm3D}
\end{equation}
Ergo, the first effect of the extension from the noncommutative plane (2D) to the noncommutative space (3D) can be seen that the  commutator in the plane is non-zero  if $i=j$. In contrast, the commutator in the noncommutative space never vanishes.

\section{3D noncommutative charged harmonic oscillator in a uniform magnetic field}
\label{sec:2}

Our starting point is the commutative Hamiltonian for the charged isotropic harmonic oscillator presence of a uniform magnetic field. 

\begin{equation}
H_0 (x,p)= \frac{1}{2m}  \left( \vec{p} - \frac{q}{c} \vec{A}\right)^2 + \frac{1}{2} m \omega^2 \left(x^2 +y^2 +z^2 \right) 
\end{equation}
Without loss of generality, we will choose the direction of the uniform magnetic field in the $z$-direction, i.e., $\vec{B} = B \hat{k}$ yielding to $\vec{A}(\vec{x},t)= \frac{1}{2} \left( -y B \hat{i} + x B \hat{j} \right)$ in Coulomb gauge. So, our Hamiltonian $H_0(x,p)$ modifies to
\begin{multline}
H_0 (x,p) = \frac{1}{2m}  \left( \left (p_x +\frac{q B}{2c}y\right)^2 + \left (p_y -\frac{q B}{2c}x\right)^2 + p_z^2 \right) +\\+ \frac{1}{2} m \omega^2 \left(x^2 +y^2 +z^2 \right). \label{eq:h02}
\end{multline}

After expanding the Hamiltonian \eqref{eq:h02} and regrouping the terms we get
\begin{multline}
H_0(x,p)= \frac{1}{2m} \left( p_x^2+p_y^2+p_z^2\right) - \frac{1}{2} \omega_c L_z + \\+\frac{1}{2}m\tilde{\omega}^2 \left(x^2+y^2\right) +\frac{1}{2} m \omega^2 z^2,
\end{multline}
where $L_z= x p_y-y p_x$ is the $z$-component of the angular momentum operator, $\omega_c= \frac{qB}{mc}$ the cyclotron frequency, and $\tilde{\omega}^2 = \omega^2+ \frac{\omega_c^2}{4}$ is the modified frequency of the harmonic oscillator in the $xy$-plane. Hence, the problem turns into the problem of an anisotropic harmonic oscillator.  From equation \eqref{eq:wm-normal}, we know that the Weyl-Moyal product can be turned into a standard product by substituting commutative $x$ and $p$ by the noncommutative operators $\hat{x}$ and $\hat{p}$, so let us first consider the Hamiltonian $H_0(\hat{x}, \hat{p})$. 
\begin{multline}
H_0(\hat{x},\hat{p})= \frac{1}{2m} \left( \hat{p}_x^2+\hat{p}_y^2+\hat{p}_z^2\right) - \frac{1}{2} \omega_c \hat{L}_z +\\+ \frac{1}{2}m\tilde{\omega}^2 \left(\hat{x}^2+\hat{y}^2\right) +\frac{1}{2} m \omega^2 \hat{z}^2 \label{eq:HNC}
\end{multline}

All noncommutative operators in the noncommutative phase-space (3D) can be stated explicitly  using \eqref{eq:xnc} and 
\eqref{eq:pnc}together with \eqref{eq:etaij} and \eqref{eq:thetaij}, respectively. 
\begin{eqnarray}
\hat{x} &=& \alpha x - \frac{\theta}{2 \alpha \hbar}  p_y +\frac{\theta}{2 \alpha \hbar}  p_z \label{eq:xNC}\\
\hat{y} &=& \alpha y - \frac{\theta}{2 \alpha \hbar}  p_z +\frac{\theta}{2 \alpha \hbar}  p_x \label{eq:yNC}\\
\hat{z} &=&\alpha z - \frac{\theta}{2 \alpha \hbar}  p_x +\frac{\theta}{2 \alpha \hbar}  p_y \label{eq:zNC}\\
\hat{p}_{x} &=& \alpha {p}_x + \frac{\eta}{2 \alpha \hbar}  y -\frac{\eta}{2 \alpha \hbar}  z \label{eq:pxNC}\\
\hat{p}_{y} &=& \alpha {p}_y + \frac{\eta}{2 \alpha \hbar}  z -\frac{\eta}{2 \alpha \hbar}  x \label{eq:pyNC} \\
\hat{p}_{z} &=& \alpha {p}_z  \frac{\eta}{2 \alpha \hbar}  x -\frac{\eta}{2 \alpha \hbar}  y \label{eq:pzNC}
\end{eqnarray} 

Based on the position and momentum operators defined in  equations  \eqref{eq:xNC}-\eqref{eq:pzNC}, we can construct all other operators needed in this calculation.

As a consequence, the noncommutative angular momentum operator $\hat{L}_z$ can be stated explicitly as following

\begin{multline}
\hat{L}_z = \hat{x} \hat{p}_y -\hat{y} \hat{p}_x = \alpha^2 L_z + \frac{\theta}{2\hbar} \left(-p_x^2 -p_y^2 +p_x p_z + p_y p_z\right) + \\+ \frac{\eta}{2\hbar} \left(-x^2-y^2+x z+ y z\right) + \frac{\theta \eta}{4\alpha^2 \hbar^2} \left(L_x + L_y+L_z\right). \label{eq:LzNc}
\end{multline}

Furthermore, the  sum of the squares of the components of the noncommutative momentum operator $\hat{p}_x^2+\hat{p}_y^2+\hat{p}_z^2$ becomes
\begin{multline}
\hat{p}_x^2+\hat{p}_y^2+\hat{p}_z^2 = \alpha^2 \left(p_x^2+p_y^2+p_z^2\right) - \frac{\eta}{\hbar} \left( L_x + L_y +L_z\right) +\\+ \frac{\eta^2}{2 \alpha^2 \hbar^2} \left( x^2 -x y+ y^2 -x z - y z + z^2 \right),\label{eq:psquareNC}
\end{multline}

and the sum of the squares of the $x$ and $y$ components of the noncommutative squared position operator $\hat{x}^2+\hat{y}^2$ is
\begin{multline}
\hat{x}^2+\hat{y}^2 = \alpha^2 \left(x^2 + y^2\right) + \frac{\theta}{\hbar} \left( -L_z + (x-y) p_z\right) + \\+ \frac{\theta^2}{4 \alpha^2 \hbar^2} \left(p_x^2 +p_y^2 +2 p_z^2-2 p_x p_z - 2 p_y p_z \right),\label{eq:xysquareNC}
\end{multline}
and finally square of the $z$ component of the noncommutative position operator $\hat{z}^2$ yields to
\begin{equation}
\hat{z}^2 = \alpha^2 z^2 + \frac{\theta}{\hbar} z \left(p_y-p_x\right) + \frac{\theta^2}{4 \alpha^2 \hbar^2} \left(p_x-p_y\right)^2. \label{eq:zsquareNC}
\end{equation}
Substituting \eqref{eq:LzNc}-\eqref{eq:zsquareNC} into \eqref{eq:HNC} gives the noncommutative Hamiltonian in the commutative algebra. After regrouping and summarizing all terms, we get the expansion of noncommutative Hamiltonian in the commutative space with respect to the noncommutativity parameters $\theta$ and $\eta$ as

\begin{multline}
H_0(\hat{x}, \hat{p}) = \alpha^2 H_0(x,p) + \frac{\eta}{\hbar} H_\eta(x,p) + \frac{\theta}{\hbar} H_\theta(x,p)+ \\+ \frac{\eta \theta}{\hbar^2} H_{\eta \theta} 	(x,p) +\frac{\eta^2}{\hbar^2} H_{\eta^2}(x,p) + \frac{\theta^2}{\hbar^2} H_{\theta^2}(x,p), \label{eq:NChamiltoniancomplete}
\end{multline}

with

\begin{eqnarray}
	H_\eta &=&-\frac{1}{2m}\left( L_x + L_y +L_z\right)-  \nonumber \\
	& & -\frac{1}{4} \omega_c\left(-x^2-y^2+x z+ y z\right), \\
	H_\theta &=&  -\frac{1}{4} \omega_c \left(-p_x^2 -p_y^2 +p_x p_z + p_y p_z\right)+ \nonumber\\
	&& + \frac{1}{2} m \tilde{\omega}^2 \bigl( -L_z + (x-y) p_z\bigr)+ \nonumber \\&&+\frac{1}{2} m \omega^2 z \left(p_y-p_x\right),\\
	H_{\eta\theta} &=&   \frac{\omega_c}{8\alpha^2 } \left(L_x + L_y+L_z\right), \\
	H_{\eta^2}&=& \frac{1}{4 m \alpha^2 } \left( x^2 -x y+ y^2 -x z - y z + z^2 \right),\\
	H_{\theta^2}&=& \frac{1}{4 \alpha^2 } \Biggl[\frac{1}{2} m \tilde{\omega}^2\left( p_x^2 +p_y^2 +2 p_z^2-2 p_x p_z - 2 p_y p_z\right) \nonumber \\ && +\frac{1}{2} m \omega^2\left(p_x-p_y\right)^2\Biggr].
\end{eqnarray}
Obviously, for $\alpha=1, \theta=\eta=0$ we return to the well known commutative case.
\section{Perturbative approach}

\label{sec:perturbation}

According to Harko et al. \cite{harko2019energy}, the contribution of the second-order terms $\eta^2$, $\theta^2$, and $\eta \theta$ can be considered as small compared to the terms in $\eta$ and $\theta$ in the low energy limit. Consequently, we can determine the effect of the noncommutativity on the binding energy by employing first-order perturbation theory. 

To determine the impact of noncommutativity on the energy levels of a charged harmonic oscillator in 3D in the presence of a uniform magnetic field, we  first have to revisit the well-known commutative case. The Hamiltonian in the commutative case in cylindrical coordinates is then given as
\begin{multline}
H_0(x,p)= -\frac{\hbar^2}{2m} \left( \frac{1}{\rho} \frac{\partial}{\partial \rho}\left(\rho \frac{\partial}{\partial \rho}\right)+ \frac{1}{\rho^2} \frac{\partial^2}{\partial \varphi^2}\right) - \frac{1}{2} \omega_c \frac{\hbar}{i}\frac{\partial}{\partial \varphi} +\\ + \frac{1}{2}m \left( \omega^2 + \frac{\omega_c^2}{4}\right) \rho^2 - \frac{\hbar^2}{2m}  \frac{\partial^2}{\partial z^2}+ \frac{1}{2} m \omega^2 z^2,
\end{multline}
where $x=\rho \cos \varphi$, $y= \rho \sin \varphi$ consequently $\rho^2= x^2+y^2$, $L_z= \frac{\hbar}{i} \frac{\partial}{\partial \varphi}$, and $p_\rho^2 = -\hbar^2 \left( \frac{1}{\rho} \frac{\partial}{\partial \rho} \left( \rho \frac{\partial}{\partial \rho} \right)\right)$. 
With $\tilde{\omega}^2 =  \omega^2 + \frac{\omega_c^2}{4}$ we get

\begin{multline}
H_0(x,p)= -\frac{\hbar^2}{2m} \left( \frac{1}{\rho} \frac{\partial}{\partial \rho}\left(\rho \frac{\partial}{\partial \rho}\right)+ \frac{1}{\rho^2} \frac{\partial^2}{\partial \varphi^2}\right)  - \frac{1}{2} \omega_c \frac{\hbar}{i}\frac{\partial}{\partial \varphi} +\\ + \frac{1}{2}m \tilde{\omega}^2 \rho^2 - \frac{\hbar^2}{2m}  \frac{\partial^2}{\partial z^2}+ \frac{1}{2} m \omega^2 z^2. 
\end{multline}

In cylindrical coordinates, the time-independent Schr\"odinger equation for a particle in an isotropic harmonic oscillator in the presence of a uniform magnetic field can be solved by separation of variables as

\begin{equation}
	\psi_{n_\rho, \mu, n_z}(x) = \chi(\rho) e^{i \mu \varphi} \zeta(z).
\end{equation}

After substitution into the time independent Schr\"odinger equation we get the eigenfunction as:
\begin{eqnarray}
	\zeta(z) &=& \frac{1}{\sqrt{2^n n!}} \left(\frac{m\omega}{\pi \hbar}\right)^{1/4} e^{-\frac{m\omega z}{2 \hbar}} H_{n_z}\left(\sqrt{\frac{m \omega}{\hbar}}z \right)\\
	\chi(\rho)&=& A_1 \left(\sqrt{2}\rho\right)^{|\mu|} e^{- \frac{m \tilde{\omega} \rho^2}{2\hbar}} U\left(-n_\rho, 1+|\mu|, \frac{m \tilde{\omega} \rho^2}{2\hbar}\right)+ \nonumber\\
	&& + A_2 \left(\sqrt{2}\rho\right)^{|\mu|} e^{- \frac{m \tilde{\omega} \rho^2}{2\hbar}} L_{n_\rho}^{|\mu|}\left(-n_\rho, \frac{m \tilde{\omega} \rho^2}{2\hbar}\right), 
\end{eqnarray}	
where $H_n(x)$ denotes the Hermite Polynomials, $U(n,m,x)$ the confluent hypergeometric function of second kind, and $L_n^\alpha(m,x)$ the generalized Laguerre Polynomial \cite{abramowitz1999ia}. The corresponding eigenvalues are given as:
\begin{equation}
	E_{n_\rho, \mu, n_z} =\hbar \tilde{\omega} (2 n_\rho+ |\mu|+1)+ \frac{1}{2}\hbar \omega_c \mu + \hbar \omega\left(n_z +\frac{1}{2}\right). \label{eq:en0}
\end{equation}

The corrections to the binding energy for weak noncommutativity in first-order perturbation theory are then according to \eqref{eq:NChamiltoniancomplete} given as

\begin{multline}
\Delta E^{(1)}_{n_\rho, \mu, n_z} = \frac{\eta}{\hbar}\braketm{n_\rho,\mu,n_z}{H_\eta}{n_\rho,\mu,n_z}+\\+\frac{\theta}{\hbar}\braketm{n_\rho,\mu,n_z}{H_\theta}{n_\rho,\mu,n_z}
\end{multline}

Due to the symmetry of the problem, all following matrix elements vanish:
\begin{multline*}
\braketm{n_\rho,\mu,n_z}{L_x}{n_\rho,\mu,n_z}=\braketm{n_\rho,\mu,n_z}{L_y}{n_\rho,\mu,n_z} = \\= \braketm{n_\rho,\mu,n_z}{xz}{n_\rho,\mu,n_z}= \braketm{n_\rho,\mu,n_z}{yz}{n_\rho,\mu,n_z} = \\=\braketm{n_\rho,\mu,n_z}{p_x p_z}{n_\rho,\mu,n_z}=\braketm{n_\rho,\mu,n_z}{p_y p_z}{n_\rho,\mu,n_z}=\\=
\braketm{n_\rho,\mu,n_z}{y p_z}{n_\rho,\mu,n_z}=\braketm{n_\rho,\mu,n_z}{x p_z}{n_\rho,\mu,n_z}=\\=
\braketm{n_\rho,\mu,n_z}{p_x}{n_\rho,\mu,n_z}=\braketm{n_\rho,\mu,n_z}{p_y}{n_\rho,\mu,n_z}=0
\end{multline*}
So, the only matrix elements that are non-vanishing are
\begin{multline}
\Delta E^{(1)}_{n_\rho, \mu, n_z} = \frac{\eta}{\hbar}\braketm{n_\rho,\mu,n_z}{-\frac{1}{2m}L_z- \frac{\omega_c}{4} \rho^2}{n_\rho,\mu,n_z}+\\+\frac{\theta}{\hbar}\braketm{n_\rho,\mu,n_z}{-\frac{\omega_c}{4} p_\rho^2 - \frac{1}{2} m \tilde{\omega}^2 L_z}{n_\rho,\mu,n_z}.
\end{multline}

With the help of \cite{SRIVASTAVA20031131,mavromatis1990interesting} the lengthy integrals can be solved in closed form, and we get for the first-order corrections in $\eta$

\begin{equation}
\Delta E_\eta^{(1)} = -\frac{\eta |\mu|}{2m}-\frac{\eta\omega_c}{4 m \tilde{\omega}} \left( 2 n_\rho+|\mu|+1\right)
\end{equation}

and $\theta$ 
\begin{equation}
\Delta E_\theta^{(1)} = - \frac{1}{2} \theta m \tilde{\omega} \left( \tilde{\omega}  - \frac{1}{2} \omega_c f( n_\rho, |\mu|)\right)
\end{equation}
with
\begin{multline}
f(n_\rho,\mu) = 2 \binomial{n_\rho+\mu}{\mu}
       - 4\mu \binomial{\mu+n_\rho+2}{n_\rho-1} - \\  
       -  \mu(1+\mu)\Biggl[ 2 \binomial{\mu+n_\rho}{n_\rho}
   +   4 \binomial{\mu+n_\rho-2}{n_\rho}+\\
   +\binomial{\mu +n_\rho+1}{n_\rho}-  \binomial{\mu +n_\rho+2}{n_\rho-1}\Biggr]. \label{eq:frhomu}
\end{multline}

A short dimensional analysis shows that $\eta$ has the dimension of $mass^2 \frac{Length^2}{Time^2}$, which corresponds to the momentum squared, and $\theta$ has the dimension of $Length^2$. So, the calculated corrections have the correct dimension of energy.

Ergo, we can summarize the results of our calculation in first-order perturbation theory. Recalling the noncommutative Hamiltonian \eqref{eq:NChamiltoniancomplete}, we see that the unperturbed energy is 

\begin{multline}
E_{n_\rho, \mu, n_z}^{(0)}=
\braketm{n_\rho,|\mu|, n_z}{\alpha^2 H_0(x,p)}{n_\rho,|\mu|,n_z} = \\ =\alpha^2 \left[ 
\hbar \tilde{\omega} (2 n_\rho+ |\mu|+1)+ \frac{1}{2}\hbar \omega_c \mu + \hbar \omega\left(n_z +\frac{1}{2}\right)
\right].\label{eq:e0final}
\end{multline}
The first-order energy corrections are 
\begin{multline}
\Delta E^{(1)}_{n_\rho, \mu, n_z}= -\frac{\eta |\mu|}{2m}-\frac{\eta\omega_c}{4 m \tilde{\omega}} \left( 2 n_\rho+|\mu|+1\right) - \\- \frac{1}{2} \theta m \tilde{\omega} \left( \tilde{\omega}  - \frac{1}{2} \omega_c f( n_\rho, |\mu|)\right)
\label{eq:final}
\end{multline}
with $f(\rho,|\mu|)$ given in \eqref{eq:frhomu}. These results hold for the situations, where $\eta \ll \hbar m \omega_c$ and $\theta \ll \frac{\hbar}{m \tilde{\omega}}$. 

\section{Discussion}
\label{sec:discussion}

Recalling one of the motivations for the development of noncommutative Quantum Mechanics was that the kinetic momentum  operators do not commute. The cyclotron frequency $\omega_c$ is directly proportional to the magnitude of the magnetic field. Therefore, let us examine the effect of the magnetic field on the energy corrections in the noncommutative phase-space.
To see the effect on the corrections clearly,  we will consider the energy correction $\Delta E^{(1)}$  normalized by $E^{(0)}$. As the energy corrections are all negative, and there is no change in sign, we will use $\left|\Delta E^{(1)}/E^{(0)}\right|$ for plotting the results.

We will employ the atomic unit system and set, therefore $\hbar$ and $m=1$.  We select arbitrarily $\omega =1$ and vary $\omega_c$ between 0.1 and 10. Based on the condition for the validity of the approximation, the values for $\theta$ and $\eta$ have to satisfy
\begin{eqnarray}
\eta &\ll& \hbar m \omega_c = \omega_c < 0.1 \text{ and }\label{etcond}\\
\theta &\ll& \frac{\hbar}{m \tilde{\omega}}= \frac{1}{\sqrt{1+\frac{\omega_c^2}{4}}} < \frac{1}{\sqrt{1+\frac{10^2}{4}}}=0.19.\label{thetcond}
\end{eqnarray}
By setting  the values $\eta=0.01$ and $\theta=0.01$, $\eta$ and $\theta$ satisfy the conditions \eqref{etcond} and \eqref{thetcond}, respectively. 

Very small values for $\theta$ need energies close to the Planck energy $E_p$, that are only available in black holes. Consequently, in this energy scale the values for $\eta$ would blowing up, and the perturbative approach would not be reasonable anymore. Therefore, as already pointed out, we select the values for $\theta$ and $\eta$ in the low energy limit. In this limit, we may have the chance to observe the effect of the corrections to the energy levels of the anharmonic oscillator due to the changing magnetic field. Therefore, the experiment has to be carried out in an environment where the change of the space-time is still observable. This indicates, that in an experimental setup, where the 3D harmonic oscillator is put in a strong magnetic field, could be method to measure the noncommutativity parameters $\theta$ and $\eta$. 

From \eqref{eq:final} it is clear that the function $f(n_\rho, |\mu|)$ plays an important role in the corrections. The possible values for $n_\rho=1,2,3$ are given in table \ref{tab1}. 

\begin{table}[h]
\begin{center}
\begin{tabular}{ccc}
$\hphantom{xx}n_\rho$ \hphantom{xx}&\hphantom{xx} $|\mu|$\hphantom{xx} &\hphantom{xx}$f(n_\rho,|\mu|)$\hphantom{xx} \\
\hline\hline
1 & 0 & 2\\
\hline
2 & 0 & 2\\
2 & 1 & -28\\
\hline
3 & 0 & 2\\
3 & 1 & -58\\
3 & 2 & -286\\
\hline
\end{tabular}
\end{center}
\caption{Values for $f(n_\rho,|\mu|)$ for $n_\rho=1,2,3$ and $|\mu|=0,..,n_\rho-1$ \label{tab1}} 
\end{table}

The energy corrections are all negative. In order to show the relation of $E^{(0)}_{n_\rho,\mu,n_z}$ and $|\Delta E^{(1)}_{n_\rho,\mu,n_z}/E^{(0)}_{n_\rho,\mu,n_z}|$ with the magnetic field, we will plot the unperturbed eigenenergy of the 3D isotropic harmonic oscillator in a uniform magnetic field as a function of $\hbar \omega_c$. Exemplarily we select $n_z=1$ and $\omega=1$ and $n_\rho=1,2,3$ and $|\mu|=0..n_\rho-1$ for the graphs of these relationships. Any other selection will not change the qualitative behavior of the system.

\begin{figure}[H]
\includegraphics[width=8.5cm]{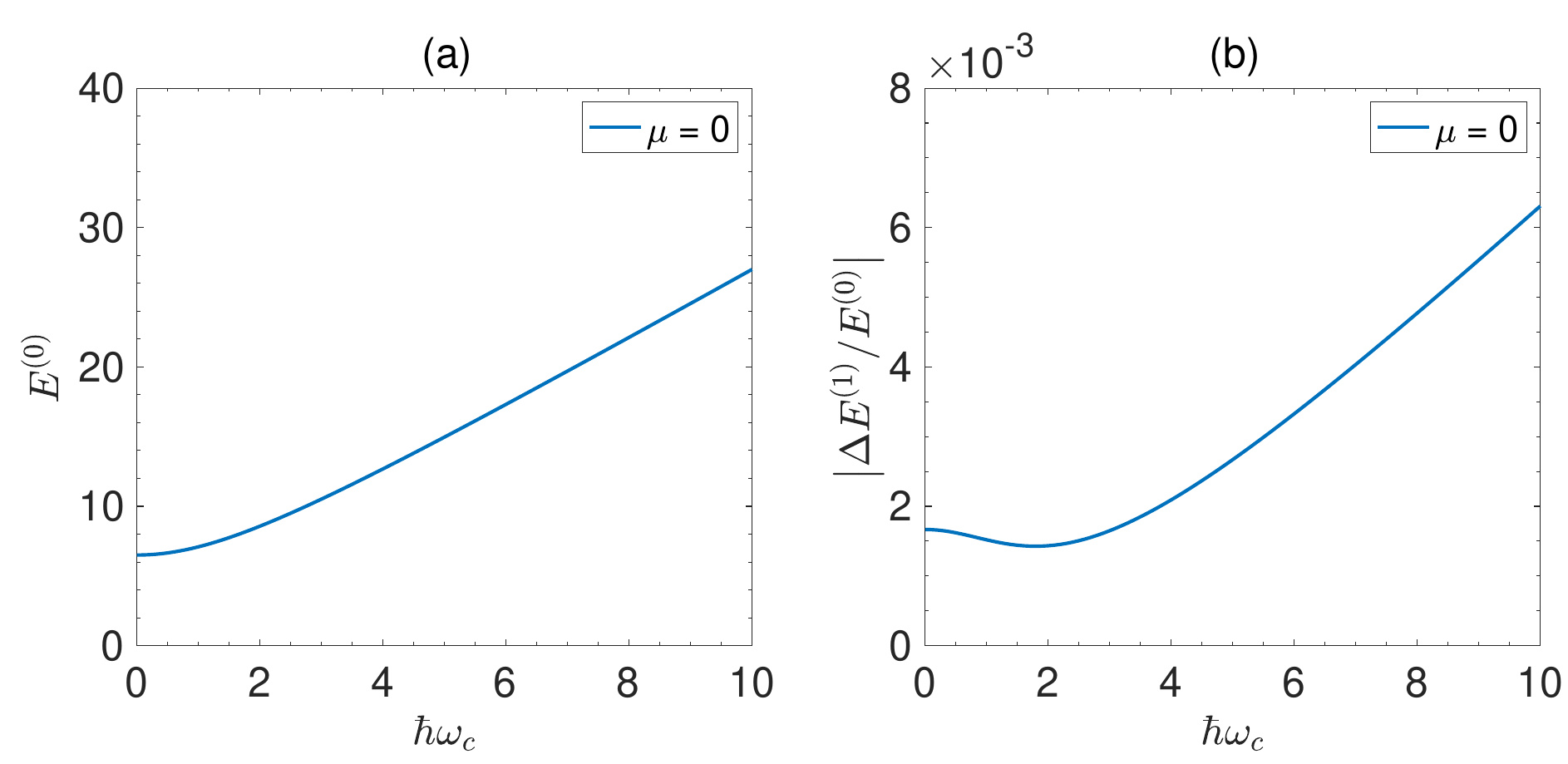} 
\caption{For the values, $n_z=1, n_\rho=1, \omega=1$, figure (a)  depicts the unperturbed 3D isotropic  harmonic oscillator in a uniform magnetic field's eigenenergy as a function of the $\hbar \omega_c$, whereas figure (b) depicts the first-order perturbation normalized by the unperturbed eigenenergy as a function of $\hbar \omega_c$. \label{fig1}}
\end{figure}

The unperturbed eigenenergy $E^{(0)}_{1,0,1}$ from \eqref{eq:e0final} for large $\omega_c$ varies asymptotically linearly with $\omega_c$. Whereas the energy correction $\Delta E^{(1)}_{1,0,1}$ varies asymptotically as  $\omega_c^2$. So $|\Delta E^{(1)}_{1,0,1}/E^{(0)}_{1,0,1}|$ will asymptotically vary $\sim \omega_c$, as depicted in figure \ref{fig1}. So, we can conclude that the magnitude of the corrections depends stronger on the magnitude of the magnetic field than the unperturbed energy of the isotropic 3D harmonic oscillator in a uniform magnetic field $E^{(0)}_{n_\rho,\mu,n_z}$. On the other hand, figure \ref{fig1} shows that for small magnetic fields, the energy corrections decrease until it reaches it local minimum at $\omega_c=1.79$ before the magnitude of the relative energy corrections starts to increase again towards its asymptotic behavior. 

\begin{figure}[H]
\includegraphics[width=8.5cm]{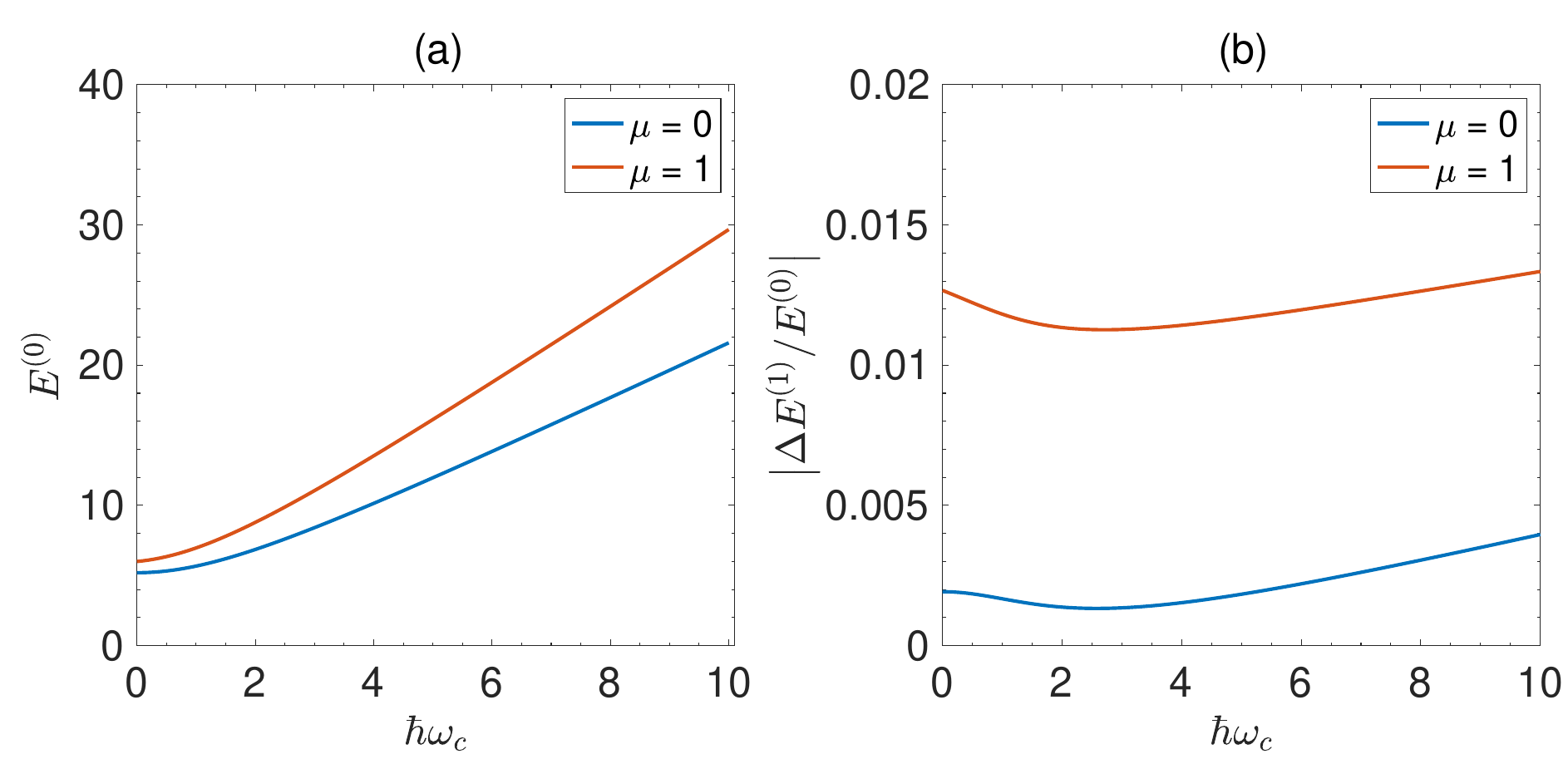}
\caption{For the values, $n_z=1, n_\rho=2, \omega=1$, figure (a)  depicts the unperturbed 3D isotropic  harmonic oscillators  in a uniform magnetic field's eigenenergy as function of the $\hbar \omega_c$ for $\mu=0$ (blue line) and $\mu=1$ (orange line), whereas figure (b) depicts the first-order perturbation normalized by the unperturbed energy as a function of $\hbar \omega_c$ for $\mu=0$ (blue line) and $\mu=1$ (orange line). \label{fig2}}
\end{figure}

In the case $n_\rho=2$, the behavior of the eigenenergies of the unperturbed isotropic 3D harmonic oscillator in a uniform magnetic field  $E^{(0)}_{2,0,1}$ and  $E^{(0)}_{2,1,1}$ and the magnitude of relative energy corrections due to noncommutativity $|\Delta E^{(1)}_{2,0,1}/E^{(0)}_{2,0,1}|$ and $|\Delta E^{(1)}_{2,1,1}/E^{(0)}_{2,1,1}|$ is qualitatively the same as in the case $n_\rho=1$. The magnitude of the relative energy correction  $|\Delta E^{(1)}_{2,0,1}/E^{(0)}_{2,0,1}|$ first decreases until $\omega_c= 2.58$, where it reaches its absolute minimum before it starts increasing again towards its asymptotic behavior. We can observe the same behavior for $|\Delta E^{(1)}_{2,1,1}/E^{(0)}_{2,1,1}|$  and get a minimum at $\omega_c=2.71$ for $\mu=1$. 
Furthermore, we can identify that for increasing magnetic Quantum number $\mu$, the magnitude of the relative corrections increases. 
\begin{figure}[H]
\includegraphics[width=8.5cm]{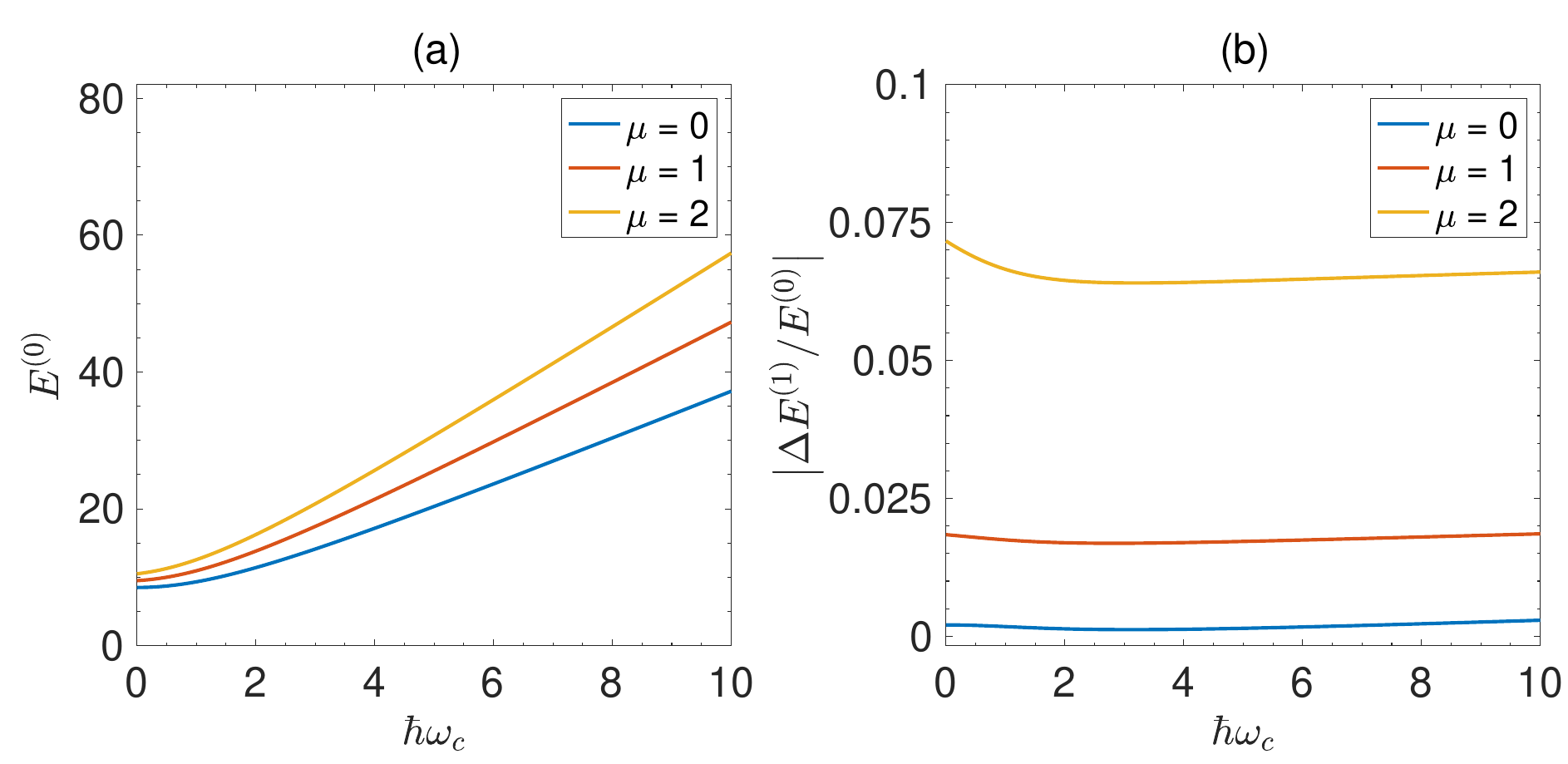} 
\caption{For the values, $n_z=1, n_\rho=3, \omega=1$, figure (a)  depicts the unperturbed 3D isotropic  harmonic oscillators  in a uniform magnetic field's eigenenergy as function of the $\hbar \omega_c$ for $\mu=0$ (blue line),  $\mu=1$ (orange line), and $\mu=2$ (yellow line), whereas figure (b) depicts the first-order perturbation normalized by the unperturbed energy as a function of $\hbar \omega_c$ for $\mu=0$ (blue line),  $\mu=1$ (orange line), and $\mu=2$ (yellow line). \label{fig3}}
\end{figure}

In the case $n_\rho=3$, the behavior of the energy of the unperturbed isotropic 3D harmonic oscillator in a uniform magnetic field and the magnitude of the relative energy corrections due to noncommutativity is qualitatively the same as in the cases $n_\rho=1$ and $n_\rho=2$. As we can see from figure \ref{fig3}, we get increasing corrections $|\Delta E^{(1)}_{3,\mu,1}/E^{(0)}_{3,\mu,1}|$ with increasing magnetic Quantum number $\mu$. The magnitude of  the relative energy corrections reach a minimum at  $\omega_c= 3.11$ for $\mu=0$, $\omega_c=2.81$ for $\mu=1$, and $\omega_c=3.20$ for $\mu=2$ before they start to increase again towards their asymptotic behavior.

Furthermore, we can  identify that the  increasing magnetic field's impact is increasing for increasing $n_\rho$ and magnetic Quantum number $\mu$. Moreover, the value for $\omega_c$ where $|\Delta E^{(1)}_{n_\rho,\mu,1}/E^{(0)}_{n_\rho,\mu,1}|$ becomes minimal increases for the constant $\mu$ and increasing $n_\rho$.

Overall, evidently, the eigenenergies and their first-order corrections strongly depend on the magnitude of the magnetic field. The relative change of the corrections to the magnetic field shows that the corrections increase faster than the eigenenergies with increasing magnetic field  for increasing magnetic Quantum numbers.

\section{Conclusion}

We studied the charged harmonic oscillator in a uniform magnetic field in the extended framework of noncommutative Quantum Mechanics in 3D. In line with this, without touching the  basic definition of the starting point of noncommutative Quantum Mechanics, namely the commutators of $[\hat{x}_i,\hat{x_j}]= i \theta_{ij}$ and   $[\hat{p}_i,\hat{p_j}]= i \eta_{ij}$ we extended the antisymmetric tensor $\lambda_{ij}$ to \eqref{eq:epsilonij3D}. This extension of the noncommutativity from the noncommutative plane to the noncommutative space gives rise to a change in the algebra of the system. The first result of this is a never-vanishing commutator $[\hat{x}_i,\hat{p_j}]$ for any combination of $i$ and $j$. Based on this algebra, we investigated the effect of the noncommutativity in 3D to the eigenenergies of the commutative system.  The Hamiltonian for the charged isotropic harmonic oscillator in a uniform magnetic field proves to be non-trivial in the noncommutative phase-space (3D). A closed solution could not be obtained in this algebra. Therefore, in the limit of weak noncommutativity, i.e., in the low energy limit, we could obtain the corrections to the eigenenergies in first-order time-independent  perturbation theory in closed form. It turns out that the corrections to the eigenenergies are negative, i.e. the eigenenergies in the noncommutative system are smaller compared to the commutative ones. To analyze the effect of the magnitude of the magnetic field on the energy corrections, we plotted the graphs of the magnitude of the relative energy corrections $|\Delta E^{(1)}_{n_\rho,\mu,n_z}/E^{(0)}_{n_\rho,\mu,n_z}|$ as a function of $\hbar \omega_c$. The analysis showed that the magnitude of the energy corrections $|\Delta E^{(1)}_{n_\rho,\mu,n_z}|$  increases asymptotically for large $\hbar \omega_c$ with $\hbar \omega_c^2$, whereas the unperturbed eigenenergies $E^{(0)}_{n_\rho,\mu,n_z}$ increase with $\hbar\omega_c$ linearly.  Ergo, the corrections to the eigenenergies increase faster with respect to $\hbar \omega_c$ than the eigenenergies themselves. This behavior could be also identified in the graphs of the relative corrections of the eigenenergies for the exemplarily selected parameters. This result suggests, that noncommuative Quantum Mechanics can be experimentally studied even in the low energy limit by employing a strong magnetic field to a 3D harmonic oscillator.

\bibliography{nc.bib}

\end{document}